# Effective masses in complex band structures, a code to extract them

Patrizio Graziosi and Neophytos Neophytou

School of Engineering, University of Warwick

Abstract

A code to extract the effective masses for the density of states and the conductivity of an arbitrary band structure is presented and the meaning of these masses for the band structure transport properties is clarified. The code, named Effective Mass Finder code (EMAF), computes the effective mass parameters for a given, generic and arbitrary, three-dimensional (3D) bandstructure of semiconductors. The code is written in MATLAB® and is released open source under the GNU GPL v3.0 license at EMAF_github and EMAF_ResearchGate.

1. **The effective mass concepts**

The presented code extracts the effective masses values that are meaningful for the charge transport, in arbitrary 3D band structures.[1]

The density-of-states (DOS) effective mass, $m_{DOS}$, is understood as the effective mass of an isotropic parabolic band that gives the same carrier density of the actual bandstructure at the band edge. The conductivity effective mass $m_{cond}$ is thought as the effective mass of an isotropic parabolic band that gives the same injection velocity in the subthreshold regime of a ballistic field effect transistor (FET), where only the band velocities are important.[2] A FET with the channel electronic bandstructure of the material under investigation is assumed. A certain Source-Drain bias is imposed, that should be relatively large as in equation A8 of reference [3]. Then the Fermi level is varied to mimic the gate bias.

2. **The DOS effective mass $m_{DOS}$**

EMAF computes the $m_{DOS}$ exploiting the facts that when the Fermi level is far from the band edge into the gap, the carrier density, per volume unit, can be expressed as

$$N = N_c e^{-\frac{E_c - E_F}{k_B T}} \qquad (1).$$

In equation (1), $E_c$ is the band edge, $E_F$ the Fermi level, $k_B$ the Boltzmann constant, $T$ the temperature, $N_c = 2\left(\frac{m_{DOS} k_B T}{2\pi \hbar^2}\right)^{3/2}$ the effective density of states in the conduction band. [4,5] For a generic numerical bandstructure $E(\mathbf{k})$ where only the conduction bands is considered, we have:

$$N = \frac{2}{(2\pi)^3} \sum_{\mathbf{k},n} f_{E_{\mathbf{k},n}} dV_{\mathbf{k}} \qquad (2)$$

where the sum runs over all the $\mathbf{k}$ points in the reciprocal unit cell, or the Brillouin zone, and all the bands, $f_{E_{\mathbf{k},n}}$ is the Fermi-Dirac distribution and $dV_{\mathbf{k}}$ is the volume element in the $\mathbf{k}$ space that usually depends only on the mesh.[6] Combining the equations (1) and (2), the effective density of states in the conduction band is obtained, then the $m_{DOS}$. The process is the same for the valence band.

### 3. The conductivity effective mass $m_{cond}$

The EMAF code computes the $m_{cond}$ from the injection velocity $v_{inj/sub}$ of the carriers in the subthreshold regime of a FET,[2]

$$m_{cond} = \frac{2k_B T}{\pi\, v_{inj/sub}^2} \qquad (3).$$

A FET with the channel electronic bandstructure of the material under investigation is assumed, working separately for the conduction bands and the valence bands. A certain Source-Drain bias is imposed, that should be relatively large,[3] 1.5 eV have been used. Then the Fermi level is varied to mimic the gate bias application. The FET current density $I_{FET}$ is computed from the difference between the Source and Drain currents for each band $n$:

$$I_{FET} = \sum_n (I_{S,n} - I_{D,n}) \qquad (4)$$

where the Source and Drain currents $I_{S,n}$ and $I_{D,n}$ are calculated as

$$I_{S,n} = e \frac{1}{2} \frac{2}{(2\pi)^3} \sum_{k_n} f_{(E_{k_n} - E_{F,S})} v_{k_n} dV_k \qquad (5a)$$

$$I_{D,n} = e \frac{1}{2} \frac{2}{(2\pi)^3} \sum_{k_n} f_{(E_{k_n} - E_{F,D})} v_{k_n} dV_k \qquad (5b).$$

In the equations (5) $e$ is the electron charge, the ½ factor it introduced for total charge conservation, $E_{F,S}$ and $E_{F,D}$ are the Source and Drain Fermi levels respectively, whose difference corresponds to the Source-Drain applied bias, $v_{k_n}$ is the carrier band velocity, computed from the gradient of the electron dispersions. $f$ and $dV_k$ are the Fermi-Dirac distribution and the volume element as in equation (2). Finally, the injection velocity is extracted as

$$v_{inj} = \frac{I_{FET}}{e \frac{1}{2(2\pi)^3} \sum_{k_n} f_{(E_{k_n} - E_{F,S})} dV_k} \qquad (6).$$

The denominator in equation (6) is the injected charge at the Source contact. Then the $v_{inj/sub}$ is identified from the operational region of the FET and the conductivity effective mass is extracted from equation (3).

The above computation scheme is conducted along the three direction *x*, *y*, and *z*. Thus, three conductivity affective masses are extracted, $m_{cond,x}$, $m_{cond,y}$, and $m_{cond,z}$ and a final average conductivity affective mass is provided as $m_{\text{cond}} = \frac{3}{m_{\text{cond},x}^{-1} + m_{\text{cond},y}^{-1} + m_{\text{cond},z}^{-1}}$. EMAF applies this scheme separately for the conduction bands and the valence bands.

### 4. Validation and examples

4.1 Parabolic bands

Parabolic bands with known analytical expression for the effective masses allow the code validation. We constructed numerical parabolic bands with isotropic effective mass of $m_0$, and with anisotropic effective masses $m_x = m_0$, $m_y = 0.5 m_0$, and $m_z = 0.1 m_0$, where $m_0$ is the electron rest mass and compared the extracted masses with the known ones. We used a cubic mesh with spacing $dk = 0.02 \pi/a$ where $a$ is 0.5 nm. Table I shows a satisfactory agreement between the nominal values and the ones extracted by the EMAF code.

| Band structure mass | Effective DOS mass | | Effective conductivity mass | |
|---|---|---|---|---|
| | Nominal | Calculated | Nominal | Calculated |
| 1 isotropic | 1 | 0.9981 | 1 | 1.0078 |
| 1, 0.5, 0.1 | 0.3684 | 0.3677 | 0.2308 | 0.2463 (1.0078, 0.5079, 0.1085) |

**Table I:** comparison between the nominal masses (in $m_0$ units) for the DOS and the conductivity, for isotropic and anisotropic bands. The values in parenthesis for the calculated conductivity masses in the anisotropic case are the three separated values extracted along the three space directions.

The concept and usefulness of the conductivity effective mass are depicted in **Figure 1**. Figure 1a shows the injection velocity versus carrier density for the cases of isotropic band, while Figures 1b and 1c report the transport distribution function (TDF) for elastic process with acoustic phonons via Acoustic Deformation Potential (ADP) and inelastic process via non-polar Optical Deformation Potential (ODP). [1,7] We used the following scattering parameters: sound velocity $1.33 \cdot 10^3$ m/s, mass density $2 \cdot 10^3$ Kg/m$^3$, ADP 10 eV, ODP $10 \cdot 10^{10}$ eV/m, $\hbar \omega_o$ 50 meV, temperature 300 K. An excellent agreement is achieved. Especially for the anisotropic case, Figure 1c, the TDF for the isotropic parabolic band having the conductivity effective mass extracted by the code reproduces very well the average TDF along the three directions. This demonstrates that the conductivity effective mass is the correct parameter to describe the transport properties of a bandstructure.

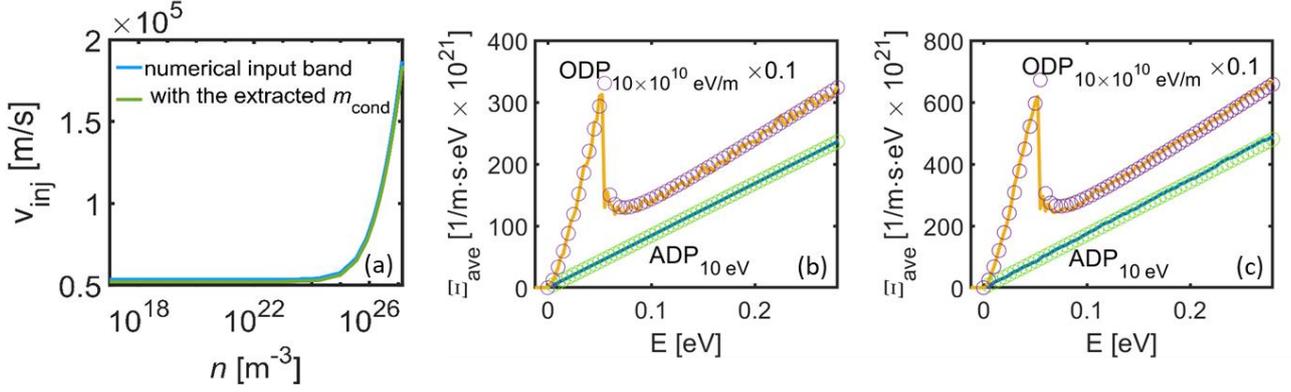

**Figure 1:** (a) injection velocity computed for the inputted numerical isotropic parabolic band (blue) and for an isotropic parabolic band with the effective mass extracted with the EMAF code. (b) and (c): average TDF $\Xi_{ave}$ for <u>isotropic</u> (b) and <u>anisotropic</u> (c) bands, the average is along the three directions. The solid lines are the TDFs as numerically extracted with the code as in ref. [7], blue for the acoustic phonons, yellow for the optical phonons. The empty circles are the analytical TDFs for isotropic parabolic bands having the $m_{cond}$ extracted with the code, green for the acoustic processes and purple for the inelastic ones. ADP is for Acoustic Deformation Potential and ODP for Optical Deformation Potential, with the scattering parameters described in the text. The match between the solid lines and the dots is remarkable.

4.2 Complex bandstructures

The case of the HfCoSb, especially of its valence band, constitutes a perfect example of complex bandstructure with many valleys and multiple minima. **Figure 2a** reports the numerical DOS for the HfCoSb, solid blue line, and the parabolic DOS that corresponds to the computed $m_{DOS}$ values, dash-dot green line. The $m_{DOS}$ concept catches the correct DOS trends at the band edge and represents an excellent and effective way to lump in a single parameter the richness of the bandstructure around the band onset.

The meaning of the $m_{cond}$ becomes clearer looking at the TDF computed for the actual numerical bandstructure and the TDF for an isotropic parabolic band with the effective mass equal to the extracted $m_{cond}$, **Figure 2b**.[1] The comparison makes sense for the isotropic scattering mechanisms only since in the parabolic approximation the distances between the $k$ points are obviously different. The blue solid line (numerical bandstructure) and the green dashed-dot line (parabolic band with $m_{\text{cond}}$) are for the ADP only mechanism. The $m_{\text{cond}}$ can clearly capture the transport properties of the complex bandstructure under elastic scattering at the band edge, when the DOS is parabolic-like. The inelastic ODP case, solid orange line (numerical bands) and purple dash-dot lines (parabolic band), presents a deviation related to the fact the final state DOS starts early to deviate from the parabolicity for phonon absorption. Thus, $m_{\text{cond}}$ can accurately describe the carrier velocity and transport properties, related to isotropic scattering with phonons, of complex bandstructures, close to the band edge. The $m_{\text{cond}}$ concepts describes the transport properties of a complex electronic bandstructure in a single parameter.

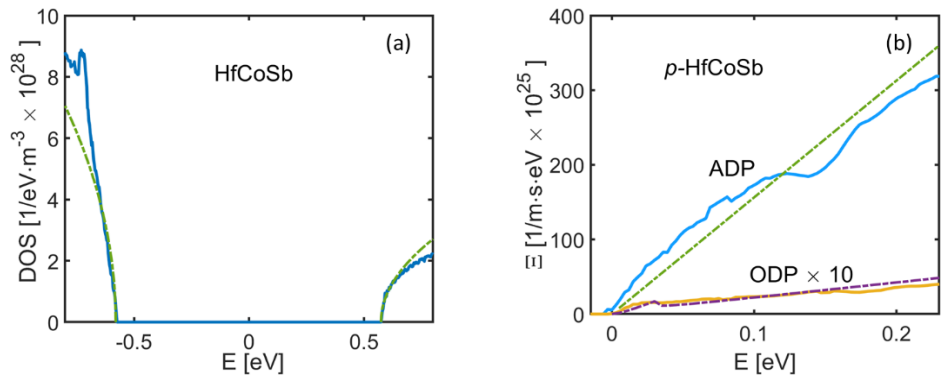

**Figure 2:** (a) DOS of HfCoSb, blue line, and DOS for parabolic bands with the isotropic $m_{\text{DOS}}$ values, green dashed lines. (b) Transport distribution function versus energy of the HfCoSb valence band for the phonon processes ADP, Acoustic Deformation Potential, and ODP, Optical Deformation Potential, blue and yellow solid lines respectively. The green and purple dashed lines represents the same quantity for a parabolic isotropic band having effective mass of $m_{\text{cond}}$.

**Acknowledgement**

This work has been funded by the Marie Skłodowska-Curie Actions under Grant Agreement No. 788465 (GENESIS—Generic semiclassical transport simulator for new generation thermoelectric materials).